\documentclass[10pt,conference]{IEEEtran}
\IEEEoverridecommandlockouts

\usepackage{cite}
\usepackage{amsmath,amssymb,amsfonts}
\usepackage{algorithmic}
\usepackage{graphicx}
\usepackage{textcomp}
\usepackage{xcolor}
\def\BibTeX{{\rm B\kern-.05em{\sc i\kern-.025em b}\kern-.08em
    T\kern-.1667em\lower.7ex\hbox{E}\kern-.125emX}}

\usepackage[T1]{fontenc}
\usepackage{newtxtext,newtxmath}
\usepackage{booktabs,url,xcolor,pifont,multirow}
\usepackage{tikz}
\usetikzlibrary{positioning,arrows.meta,fit,calc}
\usepackage[hidelinks]{hyperref}
\definecolor{blu}{HTML}{2A78D6}\definecolor{ink2}{HTML}{52514E}\definecolor{mut}{HTML}{9A9993}\definecolor{warm}{HTML}{EB6834}
\renewcommand{\arraystretch}{1.05}

\newcommand{\snap}{SNAPPerf}
\usepackage[most]{tcolorbox}
\newtcolorbox{rqanswer}{colback=black!6,colframe=black!35,boxrule=0.4pt,arc=2pt,left=3pt,right=3pt,top=1pt,bottom=1pt,before skip=3pt,after skip=3pt}

\begin{document}

\title{\snap{}: Software-Aware Tabular Foundation Model for Retraining-Free Performance Prediction of Configurable Systems}

\author{\IEEEauthorblockN{Anonymous Author(s)}}

\author{
    Amine Lbath\textsuperscript{\rm 1,2},
    Massih-Reza Amini\textsuperscript{\rm 2},
    Vadim Okun\textsuperscript{\rm 1}
    
    \thanks{Disclaimer: Certain trade names and company products are mentioned in the text or identified. In no case does such identification imply recommendation or endorsement by the National Institute of Standards and Technology (NIST), nor that they are necessarily the best available for the purpose.}\\
    \\
    \textsuperscript{\rm 1}National Institute of Standards and Technology,
    \textsuperscript{\rm 2}Université Grenoble Alpes, CNRS
}

\maketitle

\begin{abstract}
Modern software systems expose hundreds of configuration options whose combinations can substantially affect execution time, memory and energy usage. Measuring software performance resulting from a configuration set requires running a realistic workload. Since the configuration space grows exponentially, exhaustive benchmarking is infeasible. Prior work therefore trains models on a small set of measured configurations to predict the performance of the rest. Yet these models are specific to a particular software version, hardware platform, and workload, so changes require new measurements and retraining, and such changes are becoming more frequent with coding agents.

We challenge this per-system training assumption with \snap{} (Software-aware, No-retraining, Adaptive Prediction), a sample-efficient tabular foundation model specialized for configurable software. \snap{} is pretrained once on synthetic data. Each new update only requires a few measured configurations as context to predict the rest without any per-system training or hyperparameter tuning. Across 39{,}866 configurations from 12 real-world systems, spanning 6 domains and 4 performance metrics, \snap{} reduces prediction error by 38\% over the prior state of the art at the same measurement budget, and by 25\% over the best general-purpose tabular foundation model. It matches the state of the art's accuracy with 54\% fewer measurements on average. These results suggest that knowledge about software can be encoded in pretraining data rather than in per-system models, opening a new research direction in software-aware pretraining.
\end{abstract}


\section{Introduction}

Web servers, compilers, databases or video encoders expose hundreds of configuration options, and their combinations can directly influence how fast a system runs and how much memory and energy it uses. The right settings can improve performance several-fold, yet few users update configurations~\cite{xu2015knobs}, and poor settings are a frequent source of performance bugs~\cite{han2016perfbugs}. Finding good settings is hard because the number of configurations grows exponentially with the number of options, and measuring a single configuration means running a realistic workload on the entire system, which can take hours. Profiling even a simple system can consume weeks of machine time~\cite{gong2024sempl,gong2023dal}. The established approach is therefore to measure a small sample of configurations, fit a performance model to these measurements, and use this model to predict the performance of all the others~\cite{siegmund2012splc,guo2013cart,pereira2021survey}.

Prior work has proposed increasingly sophisticated learners for this task: regression trees~\cite{guo2013cart,guo2018decart}, performance-influence models~\cite{siegmund2015pim}, sparse deep networks~\cite{ha2019deepperf}, hierarchical interaction networks~\cite{cheng2023hinnperf}, and divide-and-learn ensembles~\cite{gong2023dal,gong2025dalext}. All of these methods follow the same principle: \emph{a configuration performance model is learned from measurements of the system it predicts}. Transfer and meta-learning can reuse measurements of the same system taken in another environment~\cite{valov2017transfer,jamshidi2017transfer,jamshidi2018l2s,krishna2021beetle,gong2024sempl}, but they still train a new model for each target environment.

As an example, consider a video platform that encodes uploads with the VP8 encoder (Table~\ref{tab:task}). Varying only two of its 13 options changes the encoding time of the same job elevenfold, while measuring all 2{,}736 configurations would require about 18 hours of encoding time~\cite{gong2025dalext}. To select a configuration without evaluating every candidate, engineers might measure 121 configurations, use them to train and tune a model such as DaL~\cite{gong2023dal,gong2025dalext}, and predict the remaining 2{,}615. After a software update~\cite{muhlbauer2019history} or migration to new hardware~\cite{valov2017transfer}, however, option interactions may change and the model may no longer predict performance reliably. Engineers must then collect new measurements and train and tune another model. As coding agents accelerate software development, such system updates are likely to become more frequent.

\begin{table}[h]
\caption{The prediction task on the VP8 video encoder~\cite{gong2025dalext}: each row is a configuration and its measured performance, the encoding time of the same job. The performance of every unmeasured configuration must be predicted from a small measured sample.}
\label{tab:task}
\centering\footnotesize
\setlength{\tabcolsep}{5pt}
\begin{tabular}{lcccr}
\toprule
Quality mode & Threads & Constant bitrate & \dots & Encoding time \\
\midrule
best      & 1 & no  & \dots & 60.8\,s \\
good      & 2 & yes & \dots & 22.3\,s \\
real-time & 4 & no  & \dots &  5.7\,s \\
\multicolumn{5}{l}{\emph{\dots\ 121 measured of 2{,}736 configurations}} \\
\midrule
best      & 4 & yes & \dots & \textbf{\textcolor{blu}{?}} \\
good      & 1 & no  & \dots & \textbf{\textcolor{blu}{?}} \\
real-time & 2 & yes & \dots & \textbf{\textcolor{blu}{?}} \\
\multicolumn{5}{l}{\emph{\dots\ 2{,}615 configurations to predict}} \\
\end{tabular}
\end{table}

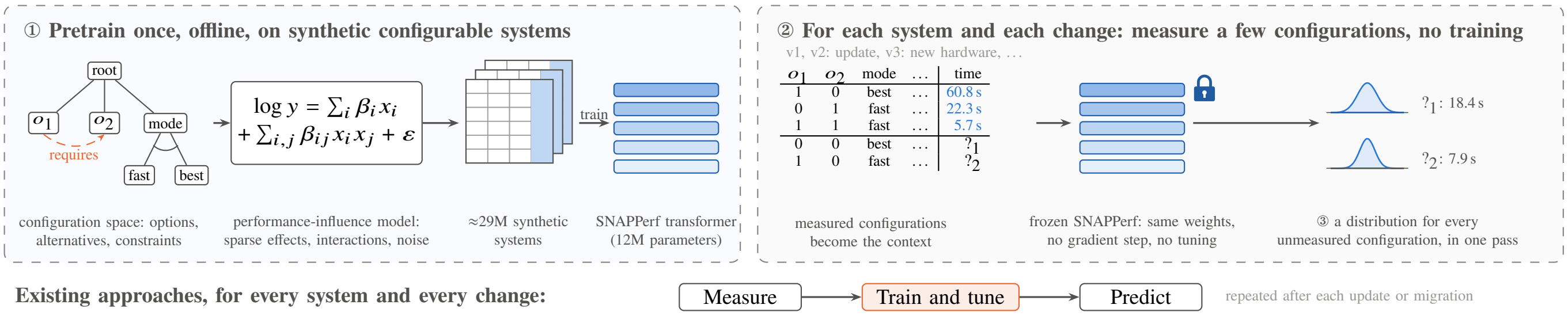
\begin{figure*}[t]
\centering
\begin{tikzpicture}[font=\scriptsize,
  panel/.style={draw=ink2!70,dashed,rounded corners=3pt},
  ptitle/.style={font=\scriptsize\bfseries,text=ink2,anchor=north west},
  lbl/.style={align=center,text=ink2,font=\tiny},
  box/.style={draw=ink2,rounded corners=1.5pt,align=center,inner sep=2pt,fill=white},
  arr/.style={-{Stealth[length=3.5pt]},ink2,semithick},
  layer/.style={draw=blu!80!black,rounded corners=1pt,minimum width=1.2cm,minimum height=0.15cm,inner sep=0pt}]

\draw[panel,fill=blu!3] (0,0) rectangle (8.45,2.95);
\node[ptitle] at (0.1,2.88) {\ding{172}\ Pretrain once, offline, on synthetic configurable systems};

\begin{scope}[every node/.style={draw=ink2,fill=white,rounded corners=1pt,inner sep=1.4pt,font=\tiny}]
  \node (r) at (1.15,2.2) {root};
  \node (a) at (0.45,1.6) {$o_1$};
  \node (b) at (1.15,1.6) {$o_2$};
  \node (g) at (1.85,1.6) {mode};
  \node (g1) at (1.55,1.0) {fast};
  \node (g2) at (2.15,1.0) {best};
\end{scope}
\draw[ink2] (r) -- (a); \draw[ink2] (r) -- (b); \draw[ink2] (r) -- (g);
\draw[ink2] (g) -- (g1); \draw[ink2] (g) -- (g2);
\draw[ink2] ($(g.south)+(-0.14,-0.14)$) arc[start angle=225,end angle=315,radius=0.2];
\draw[-{Stealth[length=2.5pt]},warm,densely dashed] (a.south) to[bend right=40] node[below,font=\tiny,text=warm,inner sep=1pt] {requires} (b.south);
\node[lbl] at (1.2,0.35) {configuration space: options,\\alternatives, constraints};

\node[box,minimum width=2.15cm,minimum height=0.95cm] (pim) at (3.7,1.6)
  {$\log y=\sum_i\beta_i x_i$\\[1pt]$+\sum_{i,j}\beta_{ij}x_ix_j+\varepsilon$};
\node[lbl] at (3.7,0.35) {performance-influence model:\\sparse effects, interactions, noise};

\foreach \dx in {0.22,0.11,0} {
  \begin{scope}[shift={(5.3+\dx,2.1+\dx)}]
    \fill[white] (0,0) rectangle (1.0,-0.96);
    \foreach \r in {0,...,5} {\foreach \c in {0,...,3} {\draw[ink2!55,line width=0.3pt] (\c*0.25,-\r*0.16) rectangle ++(0.25,-0.16);}}
    \fill[blu!30] (0.75,0) rectangle (1.0,-0.96);
    \draw[ink2,line width=0.4pt] (0,0) rectangle (1.0,-0.96);
  \end{scope}
}
\node[lbl] at (5.9,0.35) {$\approx$29M synthetic\\systems};

\foreach \i in {0,...,4} {\node[layer,fill=blu!\the\numexpr12+10*\i\relax] at (7.6,1.1+0.22*\i) {};}
\node[lbl] at (7.6,0.35) {\snap{} transformer\\(12M parameters)};

\draw[arr] (2.4,1.6) -- (pim.west);
\draw[arr] (pim.east) -- (5.22,1.6);
\draw[arr] (6.6,1.6) -- node[above,font=\tiny,text=ink2,inner sep=1pt] {train} (6.95,1.6);

\draw[panel,fill=black!2] (8.65,0) rectangle (17.9,2.95);
\node[ptitle] at (8.75,2.88) {\ding{173}\ For each system and each change: measure a few configurations, no training};
\node[font=\tiny,text=mut,anchor=west] at (8.85,2.4) {v1, v2: update, v3: new hardware, \dots};

\node[anchor=north west,inner sep=0pt] (tab) at (8.9,2.26) {\tiny\setlength{\tabcolsep}{2.6pt}\renewcommand{\arraystretch}{0.92}%
\begin{tabular}{cccc|r}
$o_1$ & $o_2$ & mode & \dots & time \\\hline
1 & 0 & best & \dots & \textcolor{blu}{60.8\,s} \\
0 & 1 & fast & \dots & \textcolor{blu}{22.3\,s} \\
1 & 1 & fast & \dots & \textcolor{blu}{5.7\,s} \\\hline
0 & 0 & best & \dots & $\mathbf{?}_1$ \\
1 & 0 & fast & \dots & $\mathbf{?}_2$ \\
\end{tabular}};
\node[lbl] at (9.95,0.35) {measured configurations\\become the context};

\foreach \i in {0,...,4} {\node[layer,fill=blu!\the\numexpr12+10*\i\relax] at (12.95,1.1+0.22*\i) {};}
\begin{scope}[shift={(13.78,1.95)}]
  \draw[blu!75!black,line width=0.9pt,line cap=round] (-0.07,0.05) -- (-0.07,0.12) arc[start angle=180,end angle=0,radius=0.07] -- (0.07,0.05);
  \fill[blu!75!black,rounded corners=0.6pt] (-0.12,-0.1) rectangle (0.12,0.07);
  \fill[white] (0,0.0) circle (0.025);
  \fill[white] (-0.011,0.0) rectangle (0.011,-0.06);
\end{scope}
\node[lbl] at (12.95,0.35) {frozen \snap{}: same weights,\\no gradient step, no tuning};

\begin{scope}[shift={(15.65,0)}]
  \foreach \yy/\m/\v in {1.72/{$\mathbf{?}_1$: 18.4\,s}/1, 1.08/{$\mathbf{?}_2$: 7.9\,s}/0.6} {
    \draw[blu,fill=blu!15,line width=0.5pt,domain=-0.42:0.42,samples=30,smooth]
      plot ({\x},{\yy+0.34*exp(-(\x*\x)/(0.025*\v))}) -- cycle;
    \draw[ink2,line width=0.3pt] (-0.47,\yy) -- (0.47,\yy);
    \node[font=\tiny,anchor=west,text=ink2] at (0.5,\yy+0.09) {\m};
  }
\end{scope}
\node[lbl] at (16.0,0.35) {\ding{174}\ a distribution for every\\unmeasured configuration, in one pass};

\draw[arr] (11.85,1.6) -- (12.25,1.6);
\draw[arr] (13.65,1.6) -- (15.1,1.6);

\node[font=\scriptsize\bfseries,text=ink2,anchor=west] at (0,-0.4) {Existing approaches, for every system and every change:};
\node[box,minimum width=1.4cm] (m) at (8.45,-0.4) {Measure};
\node[box,draw=warm,fill=warm!12,minimum width=1.8cm] (t) at (10.75,-0.4) {Train and tune};
\node[box,minimum width=1.4cm] (p) at (13.05,-0.4) {Predict};
\draw[arr] (m) -- (t); \draw[arr] (t) -- (p);
\node[font=\tiny,text=mut,anchor=west] at (13.9,-0.4) {repeated after each update or migration};
\end{tikzpicture}
\caption{Overview of \snap{}. \ding{172} A transformer-based model is pretrained once, offline, on about 29 million synthetic configurable systems. \ding{173} For any software and any update, a few measured configurations become the context of the frozen model, which \ding{174} predicts every unmeasured configuration with its uncertainty. Existing approaches instead train and tune a new model after every change (bottom).}
\label{fig:vision}
\end{figure*}

We propose predicting configuration performance from a few measurements, without retraining a model for each system. \snap{} is a single model pretrained once, independently of any target system. For a new system, the frozen model receives a few measured configurations as context and predicts the performance of all remaining configurations without updating any parameters. \snap{} is a tabular foundation model~\cite{muller2022pfn,hollmann2025tabpfn,vanbreugel2024position}, a transformer-based model that learns to predict the target column of a table from a few labeled rows by pretraining on many synthetic tables. Existing tabular foundation models are pretrained on general synthetic tables, which do not reflect the structure of configuration spaces, such as binary and numeric options, constraints between options and option interactions~\cite{siegmund2015pim}. \snap{} is therefore pretrained only on synthetic configurable systems, generated from what the literature knows about real ones. A recent survey lists few-shot learning as an open direction for configuration performance prediction~\cite{gong2025survey}. In the VP8 example, an encoder update would then require only a few new measurements, with no retraining or tuning.

Our results on the DaL benchmark~\cite{gong2025dalext} suggest that per-system training is not necessary for accurate prediction. Following the benchmark's protocol, \snap{} achieves 38\% lower prediction error than the state of the art for the same measurements and is more accurate in 57 of 60 cases. It matches the state of the art's accuracy with 54\% fewer measurements on average, and its error is 25\% lower than that of the best general-purpose tabular foundation model.

This paper makes three contributions: (i) we propose \snap{}, to our knowledge the first tabular foundation model specialized for configurable software, which is pretrained only on synthetic data and predicts performance from a few measurements without per-system retraining; (ii) we evaluate it on the DaL benchmark against prior state-of-the-art methods, five general-purpose tabular foundation models and tuned baselines; and (iii) we outline a research agenda towards software-aware pretraining.

\section{Background and Related Work}

\textbf{Configuration performance prediction.} Learning approaches range from regression trees~\cite{guo2013cart,guo2018decart} and performance-influence models~\cite{siegmund2015pim} to deep networks such as DeepPerf~\cite{ha2019deepperf} and HINNPerf~\cite{cheng2023hinnperf}, and to DaL~\cite{gong2023dal,gong2025dalext}, the current state of the art, which divides the configuration space and learns a model for each part. White-box approaches instead instrument the code~\cite{velez2021whitebox}. Complementary work studies which configurations to measure~\cite{kuhn2004fault,siegmund2012splc,sarkar2015sampling,kaltenecker2019distance,wang2026llm4perf} and shows that this choice can matter as much as the learner~\cite{kaltenecker2020interplay}. All of these approaches train and tune a new model for every system and every change, each from a fresh set of measurements. Since measurements dominate the cost, the fewer a method needs, the cheaper each change becomes.

\textbf{Transfer across environments.} When a system has already been measured in another environment, such as other hardware, transfer and meta-learning reuse these measurements to need fewer new ones~\cite{valov2017transfer,jamshidi2017transfer,jamshidi2018l2s,krishna2021beetle,gong2024sempl,dorn2026multilevel}. They still require earlier measurements of the same system and train a new model for each new environment.

\textbf{Configuration tuning.} Tuners search for the best configuration, increasingly with multi-objective and causally guided search~\cite{nair2020flash,chen2021mmo,chen2026promisetune}. \snap{} is complementary to both sampling and tuning methods, since samplers can select its input configurations and tuners can use it as their performance model.

\textbf{Foundation models in software engineering.} Large language models already replace task-specific training in software engineering, for example by reasoning over app descriptions instead of training a classifier~\cite{alecci2026reasoning}. They operate on text, however, and are less suited to numerical tables such as performance measurements~\cite{gong2025survey}. Related work also shows that a few dozen labels often suffice in software analytics~\cite{ganguly2026datalight}, which supports predicting performance from few measurements.

\textbf{Tabular foundation models.} These are transformer-based models pretrained on large and varied corpora of mostly synthetic tables, which predict the target column of a new table from a few labeled rows given as context, without further training~\cite{muller2022pfn,hollmann2025tabpfn}. They now achieve leading performance on general tabular benchmarks~\cite{grinsztajn2026tabpfn3,qu2025tabicl,zhang2025mitra,ma2025tabdpt}. Their pretraining data covers tabular problems in general. Continued pretraining on curated data can improve accuracy~\cite{garg2025realtabpfn}, but none has been tailored to software engineering.

\section{The \snap{} Model}

\snap{} consists of three components (Fig.~\ref{fig:vision}).

\textbf{\ding{172} Software-aware pretraining.} \snap{} is pretrained once, on synthetic data. Each training example is a configurable system drawn at random from a generator, which we call the \emph{prior}. Its configuration space contains binary and numeric options, groups of alternative options, and constraints between options. Its performance follows a performance-influence model~\cite{siegmund2015pim}, in which a few options and interactions dominate, numeric options can have nonlinear effects, and measurements contain noise. The prior is built from the performance-modeling literature and tuned only on separate development systems (Section~\ref{sec:results}). No system, configuration or measurement from the evaluation benchmark is used to design the prior or to train the model, so the benchmark cannot leak into pretraining. The model is a 12-layer transformer with 12M parameters, less than half the size of Nori~\cite{li2026nori}. It is trained from scratch for 38 hours on two NVIDIA H200 GPUs and learns to approximate Bayesian prediction for systems drawn from the prior~\cite{muller2022pfn}.

\textbf{\ding{173} Measurements as context.} For a given system, the measured configurations, encoded as option vectors, become the model's context. The model takes no gradient step and needs no hyperparameter search, so no part of the small sample is set aside for validation, and adapting to a change requires neither a training pipeline nor machine learning expertise.

\textbf{\ding{174} Prediction with uncertainty.} A single inference call predicts the full distribution of the performance of every unmeasured configuration. The point prediction minimizes the expected relative error, and the distribution provides an uncertainty estimate, which most existing methods lack~\cite{dorn2020uncertainty}. On our benchmark, 90\% prediction intervals contain 92\% of the true values in the median case. They can later guide which configuration to measure next.

\textbf{Use in practice.} After a software update, a hardware change or a new workload, practitioners measure a few configurations and provide them as the new context; each additional measurement adds a row. There is no per-system model to store, version or maintain.

\section{Preliminary Results}
\label{sec:results}

We evaluate \snap{} through three research questions:
\begin{itemize}\setlength{\itemsep}{0pt}\setlength{\topsep}{2pt}\setlength{\parsep}{0pt}
\item[] \textbf{RQ1:} How accurate is \snap{} compared with current state-of-the-art methods?
\item[] \textbf{RQ2:} How many measurements does \snap{} save?
\item[] \textbf{RQ3:} How does specialized pretraining compare with general-purpose tabular foundation models?
\end{itemize}

\subsection{Experimental Setup}

\textbf{Subject systems.} We use the 12 systems of the DaL benchmark~\cite{gong2025dalext}, which span databases, web servers, video encoders, compression tools, scientific solvers and image processing. Together, they provide 39{,}866 measured configurations, that is, 39{,}866 table rows, each pairing a configuration with its measured performance. Depending on the system, the predicted metric is runtime, latency, energy or maximum load.

\textbf{Development systems.} All design decisions to create our model (prior, model size, checkpoint, point estimate) were made on synthetic data or on 16 development systems from other studies~\cite{ha2019deepperf,kaltenecker2019distance,nair2020flash,krishna2021beetle}. The benchmark was used only to evaluate the final model.

\textbf{Protocol.} Following prior work~\cite{ha2019deepperf,cheng2023hinnperf,gong2023dal,gong2025dalext}, we use five measurement budgets per system, corresponding to 3\% to 22\% of all configurations. For each budget, we test on all remaining configurations and repeat each experiment 30 times with different random samples. We report the median mean relative error (MRE) over the 30 repetitions and rank methods in each case with Scott-Knott Effect Size Difference (ESD)~\cite{mittas2013sk}, which splits methods into groups whose errors differ significantly and merges groups whose difference has a negligible effect size ($\hat{A}_{12}<0.6$~\cite{vargha2000a12}). To compare \snap{} with DaL overall, we use a paired Wilcoxon signed-rank test~\cite{wilcoxon1945individual}. All tests use 99\% confidence.

\textbf{Methods.} We compare \snap{} with five general-purpose tabular foundation models: Nori~\cite{li2026nori}, TabICL~\cite{qu2025tabicl}, TabPFN~\cite{grinsztajn2026tabpfn3}, TabFM~\cite{google2026tabfm} and Mitra~2~\cite{zhang2025mitra}. At prediction time, all six models average eight predictions, each made with the options presented in a different random order or encoding. We rerun DaL and HINNPerf, the current state of the art, with their public implementations on the same samples. We also compare with four classical learners with default settings and three tuned by a cross-validated grid search within each training sample.

\begin{table}[t]
    \caption{Results over 12 systems, 5 budgets and 30 repetitions. Median MRE: median relative error (\%). Avg.\ rank: mean Scott-Knott ESD rank over the 60 cases (lower is better). Top group: cases in which a method is statistically tied for best (99\% confidence); several methods can share a case.}
    \label{tab:main}
    \centering\footnotesize
    \setlength{\tabcolsep}{4pt}
    \begin{tabular}{lccc}
    \toprule
    Method & Median MRE & Avg.\ rank & Top group (of 60) \\
    \midrule
    \multicolumn{4}{l}{\emph{Specialized foundation model [(ours)], no per-system training}} \\
    \quad \snap{} & \textbf{2.80} & \textbf{1.08} & \textbf{55} \\
    \midrule
    \multicolumn{4}{l}{\emph{General-purpose foundation models, no per-system training}} \\
    \quad Nori     & 3.80 & 2.43 & \phantom{0}9 \\
    \quad TabICL   & 4.43 & 2.72 & \phantom{0}8 \\
    \quad TabPFN   & 4.13 & 3.15 & \phantom{0}4 \\
    \quad Mitra 2  & 5.41 & 3.48 & 12 \\
    \quad TabFM    & 5.11 & 3.63 & \phantom{0}3 \\
    \midrule
    \multicolumn{4}{l}{\emph{State of the art (trained and tuned per system)}} \\
    \quad DaL      & 4.72 & 3.57 & \phantom{0}7 \\
    \quad HINNPerf & 6.41 & 4.52 & \phantom{0}6 \\
    \midrule
    \multicolumn{4}{l}{\emph{Classical learners, tuned / default}} \\
    \quad XGBoost        & 6.06\,/\,5.98 & 4.40\,/\,4.65 & 2\,/\,2 \\
    \quad Random forest  & 7.32\,/\,7.33 & 4.78\,/\,4.85 & 4\,/\,4 \\
    \quad Decision tree  & 7.93\,/\,7.69 & 4.88\,/\,5.03 & 2\,/\,3 \\
    \quad Linear regression & \phantom{0.00\,/\,}46.03 & \phantom{0.00\,/\,}6.95 & \phantom{0\,/\,}1 \\
    \end{tabular}
\end{table}

\subsection{RQ1: Accuracy}

As shown in Table~\ref{tab:main}, \snap{} has a 38\% lower median error than the prior state-of-the-art model DaL and is more accurate in 57 of the 60 system-budget cases ($p<0.001$ over all runs), with a stable gain of 37\% to 40\% across budgets. \snap{} is in the top Scott-Knott group in 55 of 60 cases, compared with 15 cases for any general-purpose foundation model, 9 for DaL or HINNPerf and 4 for any tuned classical learner. Its mean error is also lowest, at 9.8\% against 21.7\% for DaL, as it avoids the occasional large misses of other methods. In the five cases where it is not in the top group, its error is within 0.12 percentage points of the best.

\begin{rqanswer}\emph{Answer to RQ1:} \snap{} achieves 38\% lower error than DaL and is in the top group in 55 of 60 cases, wihtout requiring any training on the target system.\end{rqanswer}

\begin{figure}[t]
\centering
\includegraphics[width=\columnwidth]{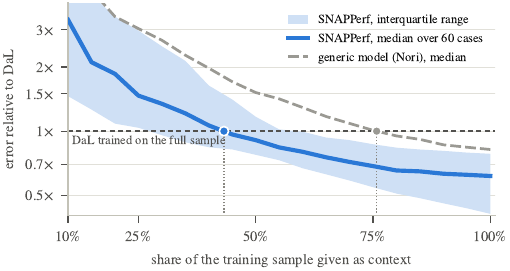}
\caption{Error given a growing share of the training sample, relative to DaL on the full sample. Below 1$\times$, the model is more accurate than DaL. Lines: median over 60 cases; band: interquartile range of \snap{}. In the median case, \snap{} matches DaL's accuracy with 43\% of the sample, and Nori, the best general-purpose model, with 76\%.}
\label{fig:savings}
\end{figure}

\subsection{RQ2: Measurement Savings}

For each case, we give \snap{} growing subsets of the training sample in steps of 5 percentage points, evaluate each on the same test set as DaL, and find the smallest subset whose median error over the 30 repetitions does not exceed DaL's error with the full sample (Fig.~\ref{fig:savings}). On average over the 60 cases, \snap{} needs 54\% fewer measured configurations than DaL, twice the 27\% saving of the best general-purpose model, Nori. The saving is 57\% for the smallest training samples and 56\% for the largest. Savings vary by system, from 10\% on HSMGP to 93\% on SQLite, where every method has a high error. The error of \snap{} keeps decreasing as it receives more measurements, indicating that it benefits from additional context at inference time rather than relying only on its pretraining.

\begin{rqanswer}\emph{Answer to RQ2:} \snap{} reaches DaL's accuracy with 54\% fewer measurements on average.\end{rqanswer}

\subsection{RQ3: Specialized vs.\ General-Purpose Pretraining}

\begin{figure}[t]
    \centering
    \includegraphics[width=\columnwidth]{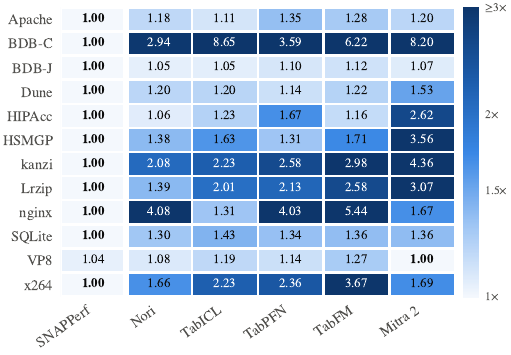}
    \caption{Error of each foundation model relative to the most accurate one in the same case, median over five budgets. Bold marks the best model per system.}
    \label{fig:spread}
\end{figure}

All six foundation models predict in the same way, by conditioning on the measured configurations, and differ mainly in their pretraining data. Among the five general-purpose models, accuracy varies widely. In a typical case, the least accurate model has twice the error of the most accurate one, and the best model differs from system to system (Fig.~\ref{fig:spread}). In contrast, \snap{} is the most accurate model on 11 of the 12 systems, and on VP8 its error is within 4\% of the best. It is more accurate than Nori, the best general-purpose model overall, in 58 of 60 cases, with a 25\% lower error in the median case. Even against the best general-purpose model of each case, selected with knowledge of the test results, \snap{} is more accurate in 54 of 60 cases. The largest improvements occur on systems where the general-purpose models perform poorly, such as BDB-C, kanzi and nginx. On the 16 development systems used for model selection, the error of \snap{} is also 32\% lower than that of Nori on average.

\begin{rqanswer}\emph{Answer to RQ3:} Specialized pretraining reduces error by 25\% compared with the best general-purpose tabular foundation model.\end{rqanswer}

\section{Future Plans}
\label{sec:plans}

The field has encoded its knowledge of configuration spaces in architectures, such as sparsity in DeepPerf or space division in DaL; \snap{} shows that it can instead be encoded in pretraining data. We plan three steps.

\textbf{Step 1: Understand the prior.} We will ablate its elements, such as interactions, constraints and noise, and extend it to coverage-based sampling designs~\cite{kaltenecker2019distance} and several metrics at once. \emph{Success criterion:} each element measurably improves accuracy on held-out systems.

\textbf{Step 2: Follow software evolution.} We will reuse measurements across releases and use predictive uncertainty to choose the next configuration to measure~\cite{muller2023pfns4bo}. \emph{Success criterion:} after an update, \snap{} needs fewer new measurements than when measuring from scratch.

\textbf{Step 3: From prediction to tuning.} Since better accuracy does not guarantee better tuning~\cite{chen2025accuracy}, we will integrate \snap{} into configuration tuners~\cite{nair2020flash,chen2026promisetune}. \emph{Success criterion:} tuners using \snap{} find better configurations for the same measurement budget.

\section{Conclusion}

\snap{} predicts the performance of configurable software from a few measurements with a model pretrained once on synthetic configurable systems. Its error is 38\% lower than the state of the art, which it matches with 54\% fewer measurements, suggesting that software knowledge belongs in pretraining data rather than in per-system models. We envision performance prediction that is available immediately after every software change, without retraining.


\bibliographystyle{IEEEtran}
\bibliography{refs}

\end{document}